\documentclass[twocolumn,,amssymb,amsmath,aps,showpacs,floatfix,nofootinbib,showpacs,balancelastpage]{revtex4}

\usepackage{hangcaption,fancybox,feynmp}
\usepackage{indentfirst}
\usepackage{graphicx}
\usepackage{epsfig,subfigure,psfrag}
\usepackage{mathrsfs}

\begin{document}

\include{newcmd}
%%%%%%%%%%%%%%%%%%%%%%%%%%%%%%%%%%%

\title{One-side Forward-backward Asymmetry in Top Quark Pair Production at CERN Large Hadron Collider }
\author{You-kai Wang$^{1}$\footnote{E-mail:wangyk@pku.edu.cn},
Bo Xiao$^{1}$\footnote{E-mail:homenature@pku.edu.cn},
 and Shou-hua
Zhu$^{1,2}$\footnote{E-mail:shzhu@pku.edu.cn} }

\affiliation{ $ ^1$ Institute of Theoretical Physics $\&$ State Key
Laboratory of Nuclear Physics and Technology, Peking University,
Beijing 100871, China \\
$ ^2$ Center for High Energy Physics, Peking University, Beijing
100871, China }

\date{\today}

\begin{abstract}

Both D0 and CDF at Tevatron reported the measurements of
forward-backward asymmetry in top pair production, which showed
possible deviation from the standard model QCD prediction. In this
paper, we explore how to examine the {\em same} higher-order QCD
effects at the more powerful Large Hadron Collider.

\end{abstract}

\pacs{14.65.Ha, 12.38.Bx}

\maketitle

The top quark is the heaviest ever known fermion and is thought to
be related to the mechanism of electroweak symmetry breaking and
physics beyond the standard model (SM). Since it was discovered more
than one decade ago, measuring its properties is one of the most
active fields.  Most of the measured properties such as mass, width,
production rate and so on are consistent with SM predictions,
however the CDF and D0 Collaborations have observed a possible
deviation on forward-backward (FB) asymmetry. At $t\bar t$ frame
$A_{\rm FB}$ is defined as \begin{equation}
\begin{array}{rl}
A_{\rm FB}&=\frac{\sigma( \Delta Y>0)-\sigma(\Delta
Y<0)}{\sigma(\Delta Y>0)+\sigma(\Delta Y<0)}, \label{originAFB}
\end{array}
\end{equation}
where $\Delta Y\equiv Y_t-Y_{\bar t}$ is the difference between
rapidity of the top and antitop quark, which is invariant under
$t\bar t$ or $p\bar{p}$ rest frame.

The measurements of CDF and D0 are
\cite{ICEHP2010Eppig,ICEHP2010Shary},
\begin{equation} \begin{array}{rl}
A_{\rm FB}^{CDF} &= 0.158\pm 0.072\pm 0.017, \ \ {\rm with}\ \   \ 5.3 \mbox{fb}^{-1};  \\
A_{\rm FB}^{D0}  &= 0.08\pm 0.04\pm 0.01, \ \ {\rm with}\ \   \ 4.3
\mbox{fb}^{-1}.
\end{array}
\end{equation}
 The measurements are consistent with
previous ones \cite{Aaltonen:2009iz,Abazov:2008,Aaltonen:2008hc}.
The corresponding SM predictions from Monte Carlo(MC) simulations
are $5.8\pm 0.9 \%$ \cite{ICEHP2010Eppig} by MCFM and $1^{+2}_{-1}
\%$ \cite{ICEHP2010Shary} by MC@NLO. Here D0's measurement and the
corresponding MC@NLO prediction can not be compared directly with SM
ones because they are not normalized by selecting efficiency. The
$A_{\rm FB}$ in the SM is calculated to be 7.8\% in Ref.
\cite{Kuhn:1998jr,Kuhn:1998kw,Antunano:2007da}, which is larger than
$5.8\pm 0.9 \%$ \cite{ICEHP2010Eppig} by MCFM. The reason is that
the denominator of $A_{\rm FB}$ in the MCFM is the cross section at
the next leading order QCD, while the leading order cross section in
Ref. \cite{Kuhn:1998jr,Kuhn:1998kw,Antunano:2007da}. Therefore they
differ by a factor of $k\sim 1.3$.

\begin{figure}[htbp]
\centerline{\hbox{
\includegraphics[height=2.5cm,width=2.5cm]
{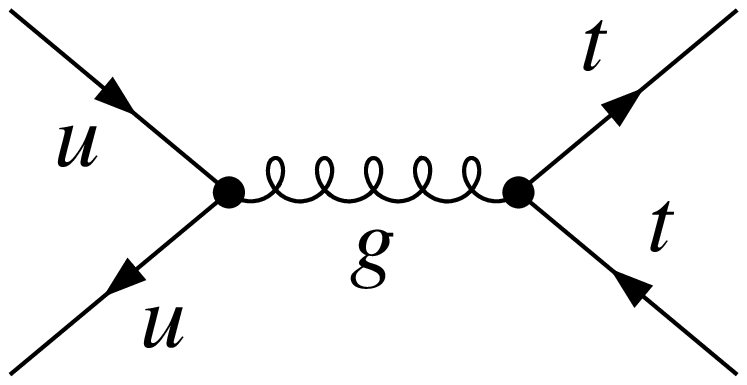}
\includegraphics[height=2.5cm,width=2.5cm]
{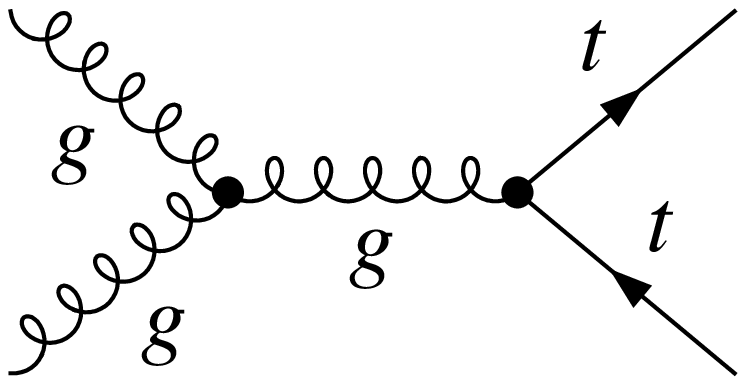}
\includegraphics[height=2.5cm,width=2.5cm]
{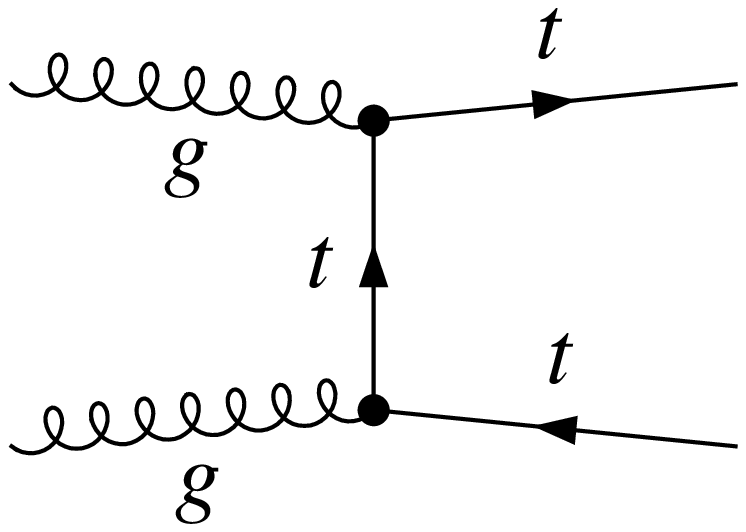} }} \caption{\label{tree} Typical
Feynman diagrams for $t\bar t$ pair production at LHC at
$O(\alpha_s^2)$. }
\end{figure}

\begin{figure}[htbp]
\centerline{\hbox{
\includegraphics[height=3.0cm,width=3.0cm]
{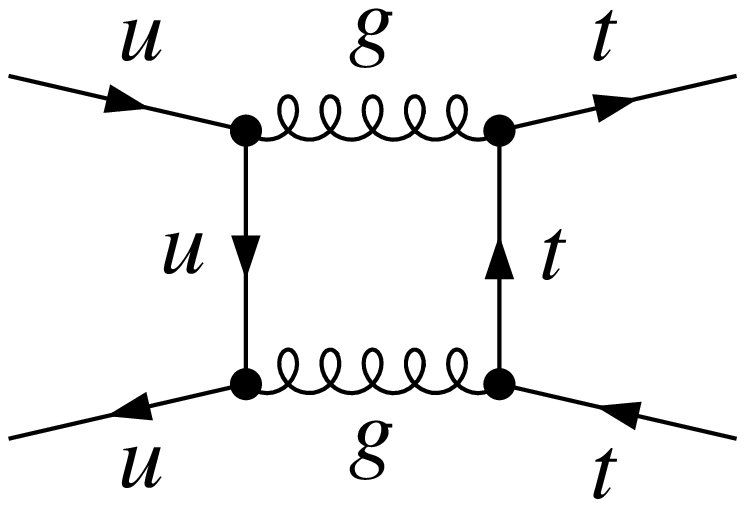}
\includegraphics[height=3.0cm,width=3.0cm]
{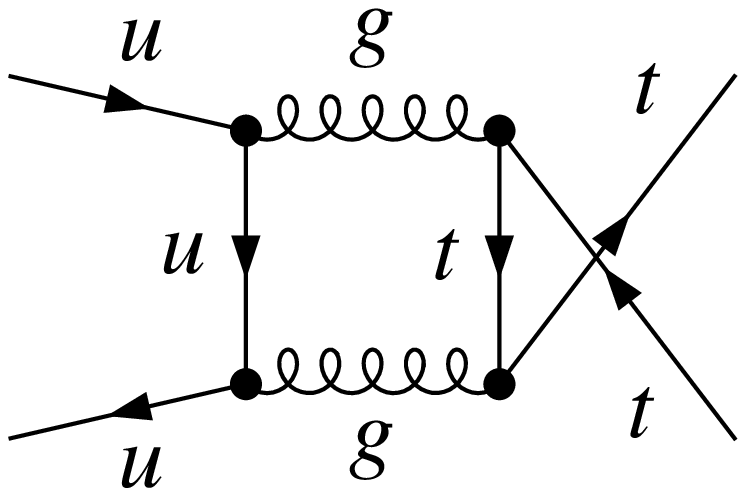} }} \caption{\label{box} Typical NLO
 virtual Feynman diagrams which contribute to asymmetric cross section. }
\end{figure}

\begin{figure}[htbp]
\centerline{\hbox{
\includegraphics[height=2.5cm,width=2.5cm]
{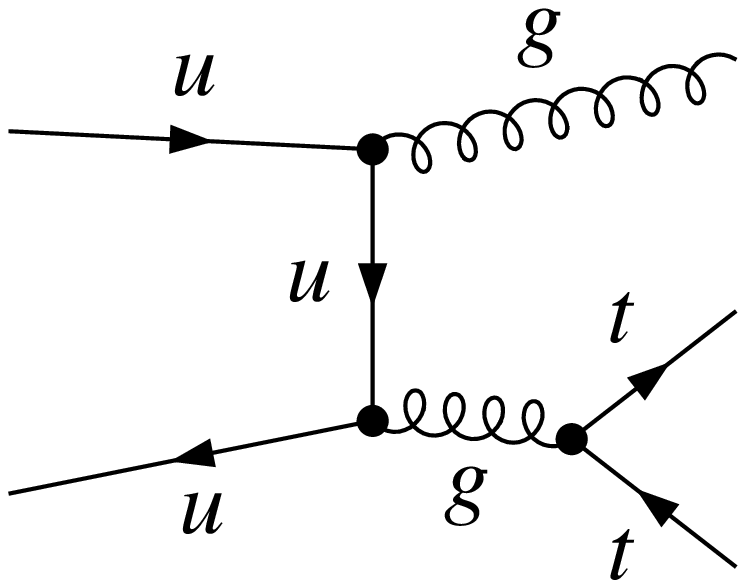}
\includegraphics[height=2.5cm,width=2.5cm]
{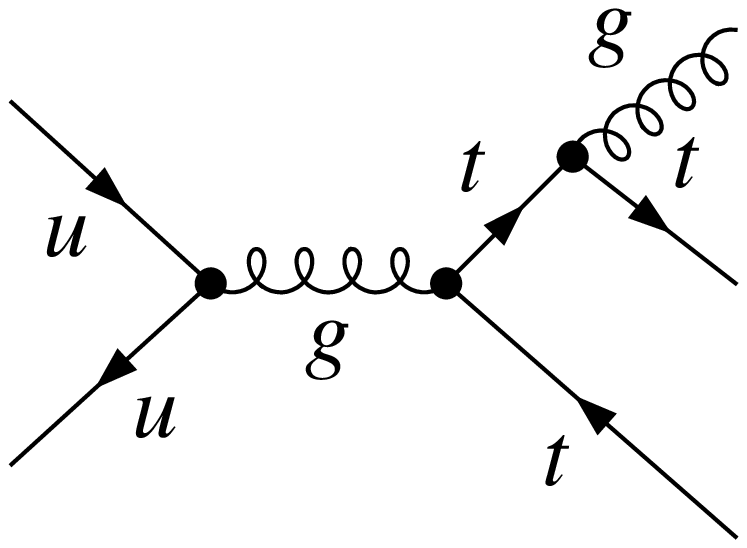} } } \caption{\label{uuttg}
Typical real gluon emission Feynman diagrams which contribute to
asymmetric cross section. }
\end{figure}

Such FB asymmetry is equivalent to the charge asymmetry provided
that $CP$ is conserved. It is strange at first glance that vector
like theory QCD can induce FB asymmetry. The fact is that such
asymmetry arising from higher-order effects, namely, the
interference between tree-level and virtual box diagrams of $t\bar
t$ production, as well as among diagrams of real processes of $q\bar
q \rightarrow t \bar t g$ (cf. Figs.~\ref{tree}-\ref{uuttg}).
Similar asymmetry of QED was noticed even 37 years ago
\cite{Berends:1973fd}.

Obviously only less than $3\sigma$ deviation is not the evidence
that the SM is failed. Though the pursuit of possible new physics
beyond the SM (BSM) implied by the deviation is exciting, the
investigation of the same inference effect at more powerful Large
Hadron Collider (LHC) is more necessary. Once the deviation is
confirmed at the LHC, the measurements may be the first BSM
signature. Unfortunately, the FB asymmetry defined at the
proton-antiproton collider Tevatron is not applicable at the  {\em
proton-proton} collider LHC, as LHC does not have the preferred
direction in the laboratory frame. In order to solve this issue, the
central charge asymmetry has been proposed
\cite{Kuhn:1998jr,Kuhn:1998kw,Antunano:2007da,
Rodrigo:2008qe,Ferrario:2008wm, Ferrario:2010hm}
\begin{equation}
A_C=\frac{\sigma_t(|Y|\leq Y_C)-\sigma_{\bar{t}}(|Y|\leq
Y_C)}{\sigma_t(|Y|\leq Y_C)+\sigma_{\bar{t}}(|Y|\leq Y_C)}.
\label{ACharge}
\end{equation}
Here $A_C$ is defined as the ratio between the difference and the
sum of the events of the top and the antitop quark in a central
region $|Y|<Y_C$ in the laboratory frame. The disadvantage of this
definition is that at the LHC, such asymmetry is quite small. The
reason is that the central region cut $|Y|<Y_C$ can not remove the
symmetric $t\bar t$ events via $gg$ fusion efficiently. In this
paper, we propose a new definition of FB asymmetry, namely the
one-side FB asymmetry $A_{\rm OFB}$, to conquer this difficulty.
$A_{\rm OFB}$ can be large and arises from the {\em same}
$O(\alpha_s^3)$ contributions which induce the observed FB asymmetry
at Tevatron. This quantity can be examined at the LHC and
cross-checked to the corresponding measurements at the Tevatron.

At the LHC, up to next-to-leading order (NLO) QCD, top pair events
can be generated through the channels $q\bar{q}\to t\bar{t}$,
$q\bar{q}\to t\bar{t}g$, $q g\to t\bar{t}q$ or $\bar{q} g\to
t\bar{t}\bar{q}$ and $g g\to t\bar{t}$ at the partonic level. Being
a proton-proton collider, LHC does not have the preferred directions
in the laboratory rest frame. However except the symmetric gluons,
the incoming partons do have preferred direction. Usually the
valence quark momentum is larger than that of the sea quark. For
example, for the process $u\bar{u}\to t\bar{t}$ (taking the momentum
of the $u$ quark as the positive $z$ direction), momentum of $u$ is
most probably larger than that of $\bar{u}$. On average, this will
induce the z direction $t\bar{t}$ total momentum in lab frame
$P_{t\bar{t}}^z>0$. So even for the $pp$ collider, $u\bar{u}\to
t\bar{t}$ can contribute an asymmetric $t\bar{t}$ distribution.
However, this asymmetry is completely canceled with the opposite
direction $\bar{u}u\to t\bar{t}$ events. If we observe only one-side
$t\bar t$ events, i.e. $P_{t\bar{t}}^z>0$, such asymmetry will be
kept. To maintain the partonic asymmetry and suppress the symmetric
events, we require a cut $|P_{t\bar{t}}^z|>P_{cut}^z$ on the z
direction top pair momentum of the final $t\bar t$ pair in the $pp$
rest frame. One may argue that determination of the momentum in beam
line direction may be problematic, especially when one neutrino is
missing when using the associated charged lepton to trigger the
top/antitop event. This issue can be solved by requiring invariant
mass of the neutrino and charged lepton just equal to that of the
$W$ boson, which is assumed to be the decay product of the top
quark. Thus $P_{t\bar{t}}^z$ is still a measurable quantity
\cite{CDFnote9724}.

The new one-side forward-backward asymmetry $A_{\rm OFB}$ can be
defined in the $pp$ rest frame as
\begin{equation}
\begin{array}{rl}
A_{\rm OFB}&=\frac{\sigma( \Delta Y>0)-\sigma(\Delta
Y<0)}{\sigma(\Delta Y>0)+\sigma(\Delta Y<0)}
|_{P_{t\bar{t}}^z>P_{\rm cut}^z,M_{t\bar{t}}>M_{\rm cut}}\\\\
&=\frac{\sigma(\Delta Y<0)-\sigma(\Delta Y>0)}{\sigma(\Delta
Y<0)+\sigma(\Delta Y>0)} |_{P_{t\bar{t}}^z<-P_{\rm cut}^z,
M_{t\bar{t}}>M_{\rm cut}} \label{AFB}
\end{array}
\end{equation}
or
\begin{equation}
A_{\rm OFB}=\frac{F_- + B_-}{F_+ +B_+}\equiv\frac{\sigma^A}{\sigma},
\label{AFB2}
\end{equation}
with
\begin{equation} F_\pm= \left. \left(\sigma( \Delta Y>0)\pm \sigma(\Delta
Y<0)\right)\right|_{P_{t\bar{t}}^z>P_{\rm cut}^z,M_{t\bar{t}}>M_{\rm
cut}}
\end{equation}
\begin{equation} B_\pm= \left. \left(\sigma(\Delta
Y<0)\pm \sigma(\Delta Y>0)\right)\right|_{P_{t\bar{t}}^z<-P_{\rm
cut}^z, M_{t\bar{t}}>M_{\rm cut}} \end{equation}

The asymmetry defined in  Eq.(\ref{AFB2}) is the same as that in
Eq.(\ref{AFB}) except the statistics are doubled. We will adopt the
asymmetry definition in Eq.(\ref{AFB2}) in the following evaluation.
The goal to apply constraint on $P_{t\bar{t}}^z$ and $M_{t\bar{t}}$
 is to exclude the symmetric $gg\to t\bar{t}$ events.
 In Eq.(\ref{AFB2}), the asymmetric cross section in the numerator arises from
$O(\alpha_s^3)$ in QCD, and the denominator is the total cross
section. Although some high order effects in $t\bar{t}$ production
have been considered, such as soft gluon resummation
\cite{Almeida:2008ug, Ahrens:2010zv} and the exclusive
next-to-leading order cross section of $t\bar{t}+jet$
production\cite{PhysRevLett.98.262002, Dittmaier:2008uj,
Melnikov:2010iu}, the exact inclusive next-to-leading order
asymmetric cross section which involves the two-loop contributions
is still unknown. For consistency, we choose the lowest order result
of total cross section at $O(\alpha_s^2)$ as a rough estimation.

The typical Feynman diagrams of $O(\alpha_s^2)$ for the denominator
in Eq.~\ref{AFB} are drawn in Fig.~\ref{tree}.

\begin{figure}[htbp]
\centerline{\hbox{
\includegraphics[height=2.5cm,width=2.5cm]
{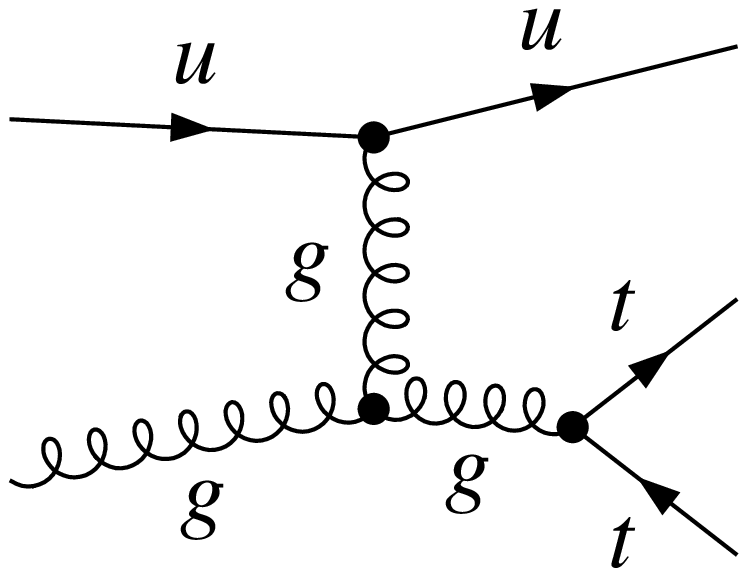}
\includegraphics[height=2.5cm,width=2.5cm]
{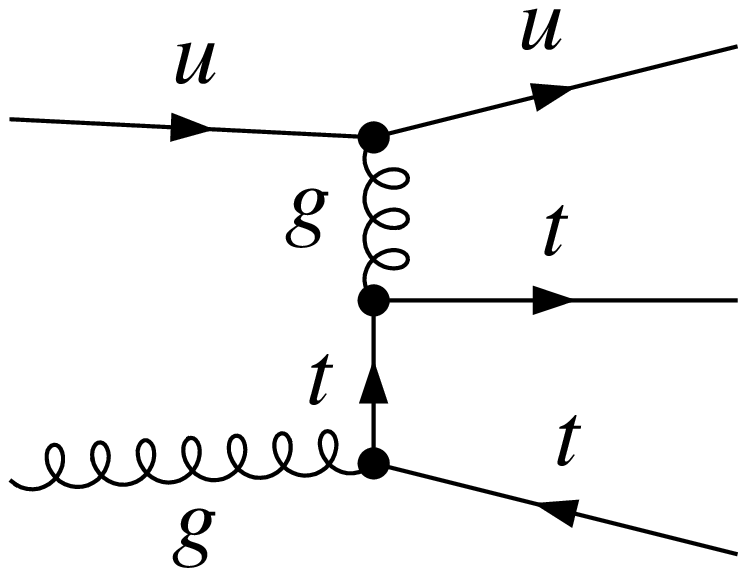}
\includegraphics[height=2.5cm,width=2.5cm]
{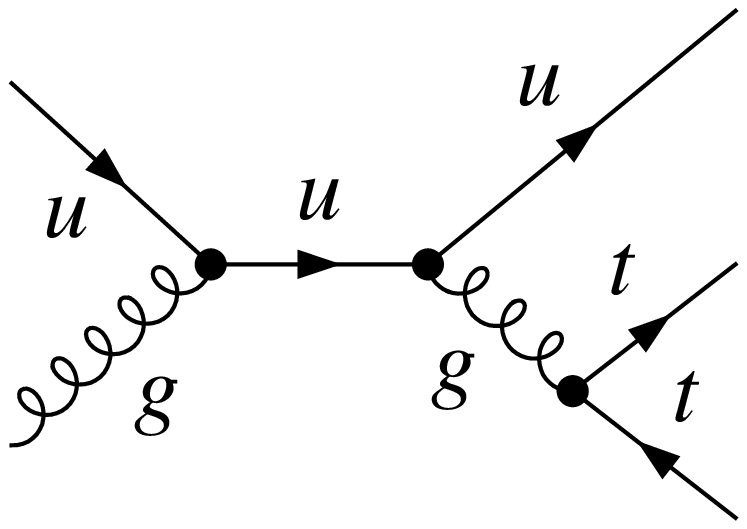} }} \centerline{(a)} \centerline{\hbox{
\includegraphics[height=2.5cm,width=2.5cm]
{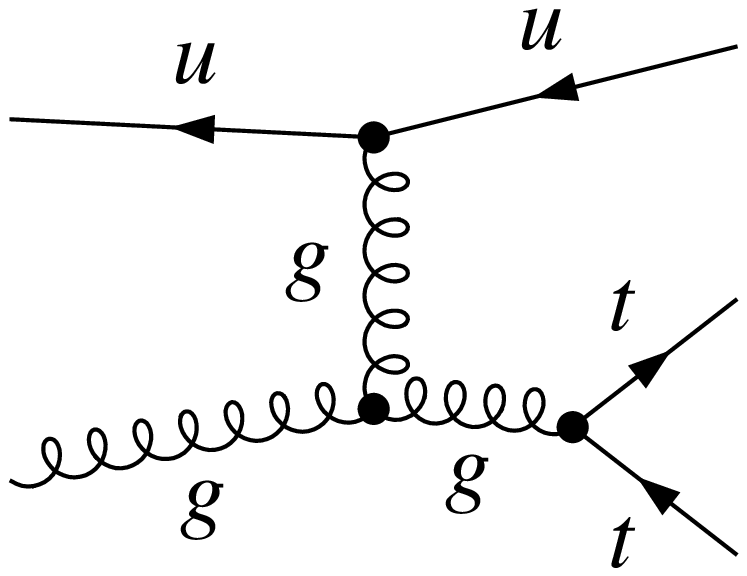}
\includegraphics[height=2.5cm,width=2.5cm]
{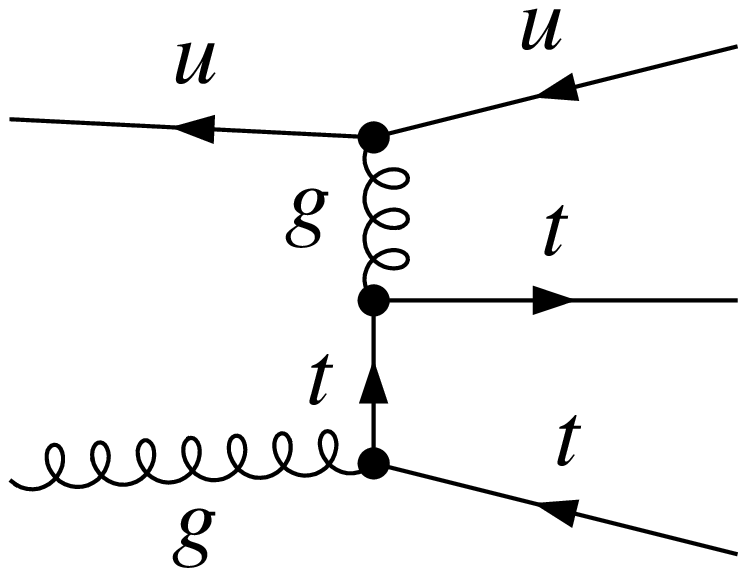}
\includegraphics[height=2.5cm,width=2.5cm]
{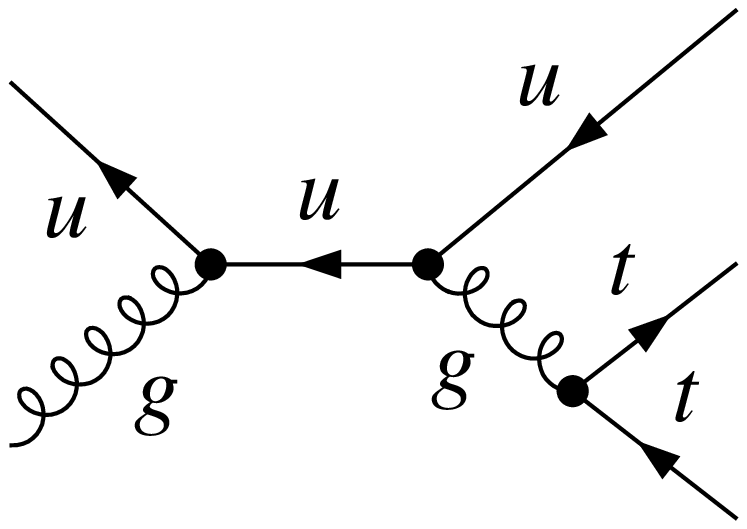} }} \centerline{(b)}
\caption{\label{ugttu} Typical Feynman diagrams of $u g \to u t\bar{t}$ (a), and $\bar{u} g \to
\bar{u} t\bar{t}$ (b).}
\end{figure}

In the SM, the leading asymmetric cross section arises  at
$O(\alpha_s^3)$. The related Feynman diagrams at partonic level can
be classified into three categories: (1) the interference among
virtual box in Fig.~\ref{box} and the leading diagrams for the
process $q\bar q \rightarrow t \bar t$ in Fig.~\ref{tree}; (2) the
interference among initial and final gluon radiation diagrams of
$q\bar q \rightarrow t \bar t g$ in Fig.~\ref{uuttg}; and (3)
contributions from diagrams of  $q g\to t\bar{t}q$ and $\bar{q} g\to
t\bar{t}\bar{q}$ in  Fig.~\ref{ugttu}.

The asymmetric cross section at the parton level was given
analytically in Ref.~\cite{Kuhn:1998kw}. However, we carry out
independent calculations \cite{Xiao:2010hm} with the help of
FeynCalc~\cite{Mertig:1990an}, FormCalc~\cite{Hahn:1998yk} and
QCDLoop~\cite{Ellis:2007qk}.

The asymmetric cross section $\sigma^A$ contributed from each of the
above three categories is UV and collinear divergences free. The
real gluon radiation of category (2) can be divided into a soft part
and a hard part by introducing a soft cut $\delta^s$
\cite{Harris:2001sx}. Soft divergences are canceled when adding the
virtual (1) and soft part to form a virtual-soft part. $\delta^s$
independence is checked by adding the virtual-soft and hard part
\cite{Xiao:2010hm}.

 In our numerical calculations, we
choose cteq6l for the leading order calculation and cteq6m for
higher-order estimations. The scales are chosen as $\mu_r=\mu_f=m_t$
and $\alpha_S(m_Z)=0.118$.

To check our Fortran code, forward-backward asymmetry at Tevatron is
recalculated independently using our code. $A_{\rm FB}$ is
calculated to be $7.1\%$, which is in good agreement with the
existing results \cite{Antunano:2007da,Bernreuther:2010ny}.

\begin{figure}[htbp]
\includegraphics[height=5cm]
{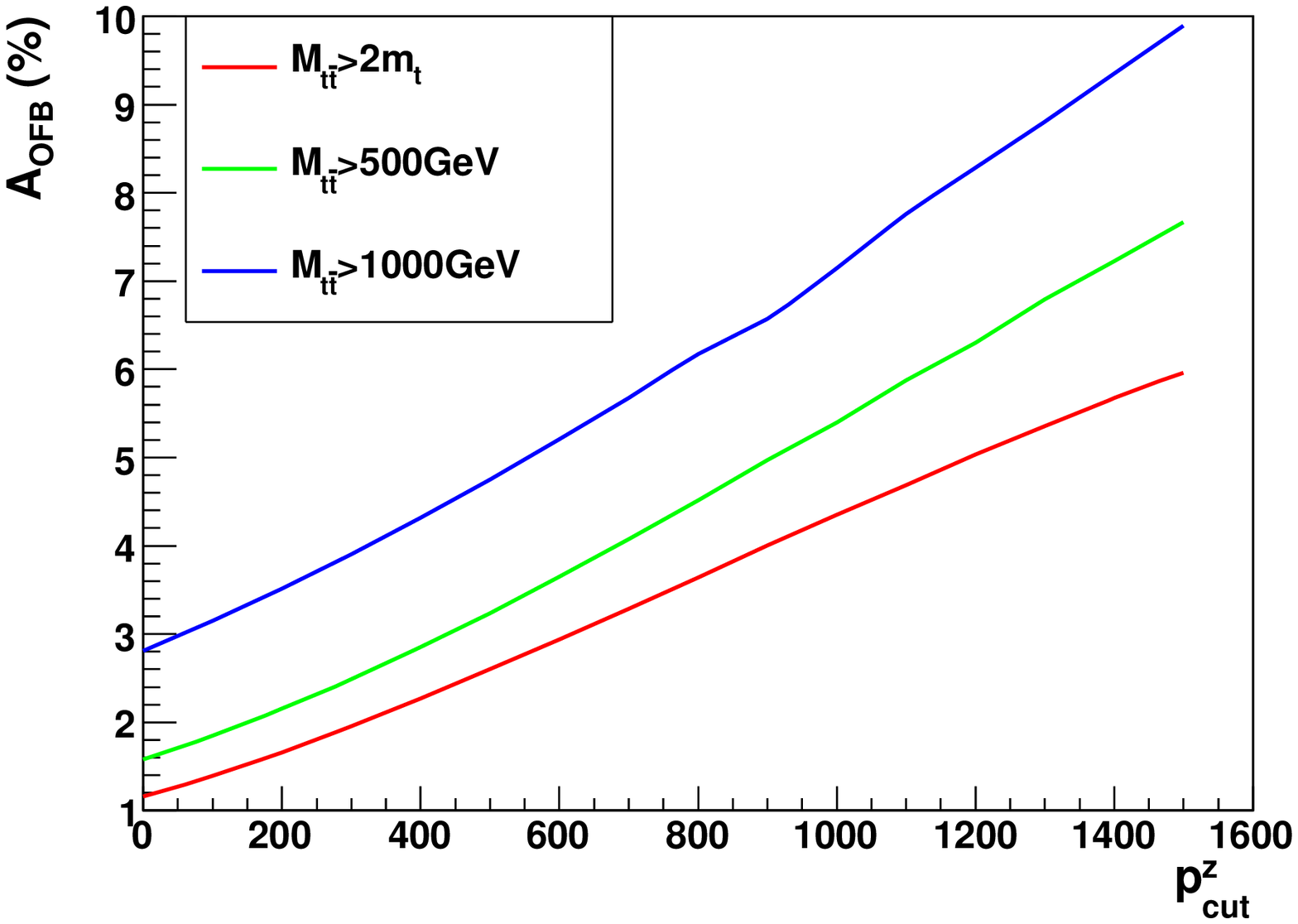}\\
\includegraphics[height=5cm]
{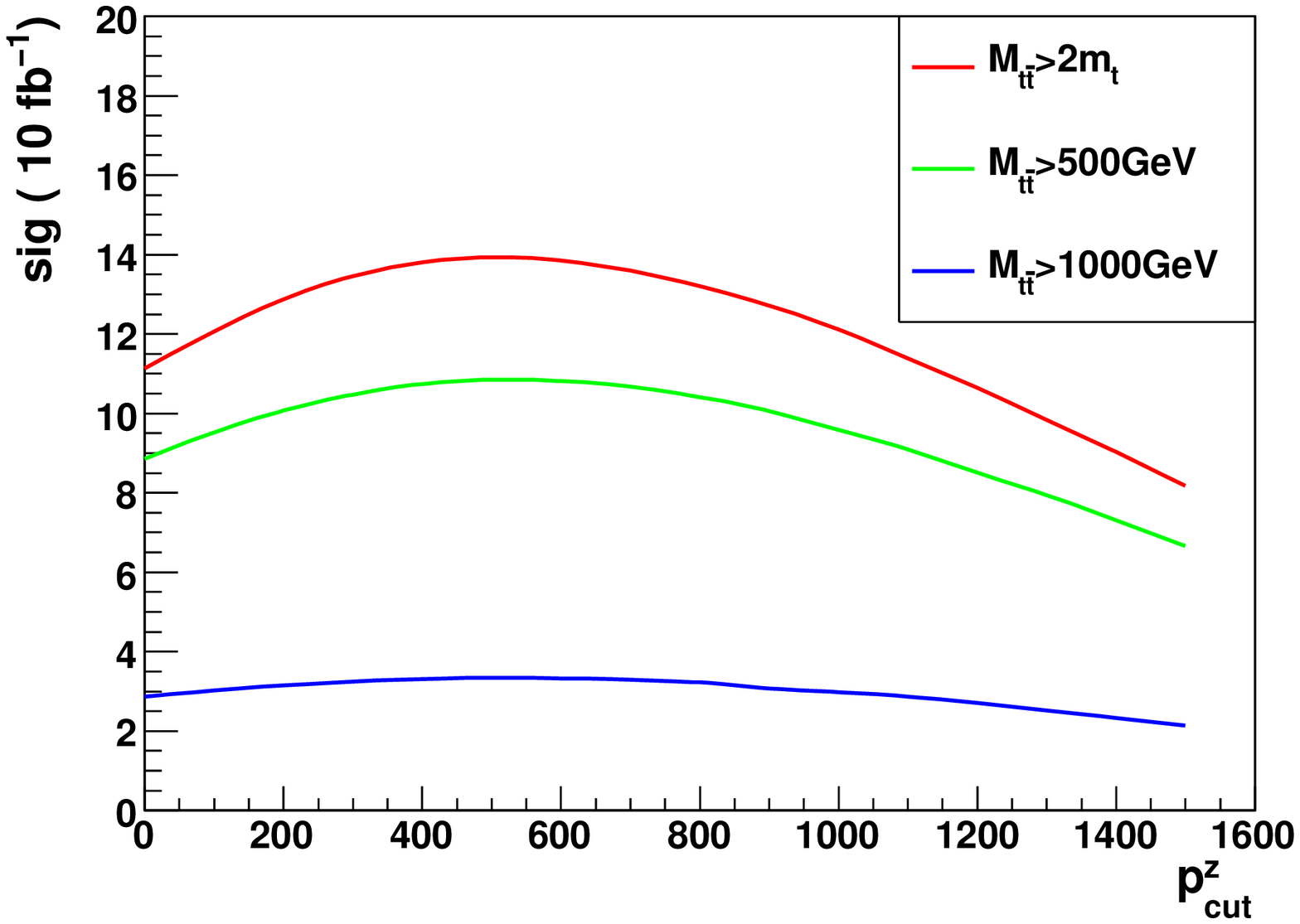} \caption{\label{Afb7TeV}
$A_{\rm OFB}$ and significance as a function of  $P_{cut}^z$ at LHC
with $\sqrt{s}=7$ TeV.}
\end{figure}

\begin{figure}[htbp]
\includegraphics[height=5cm]
{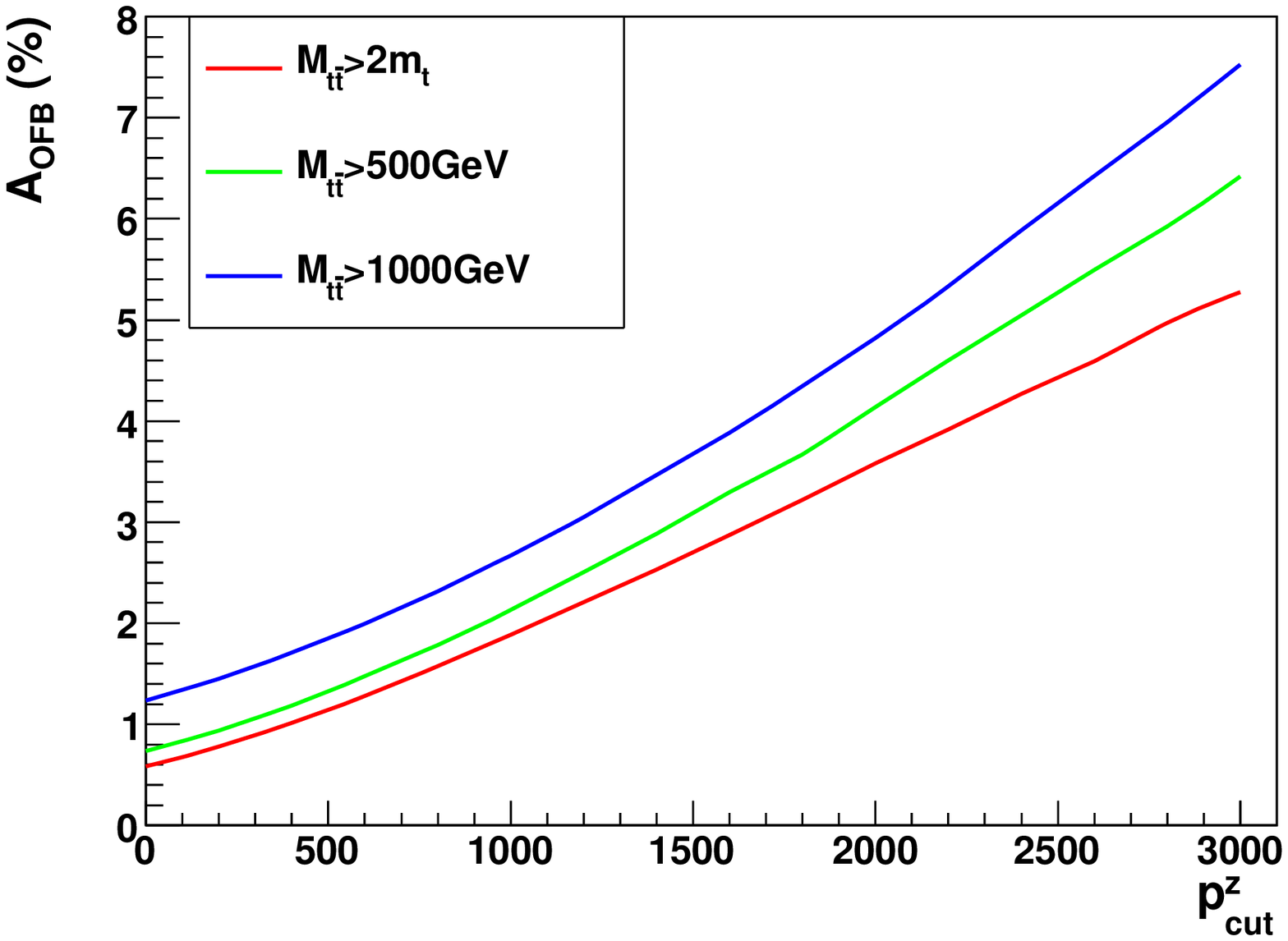}\\
\includegraphics[height=5cm]
{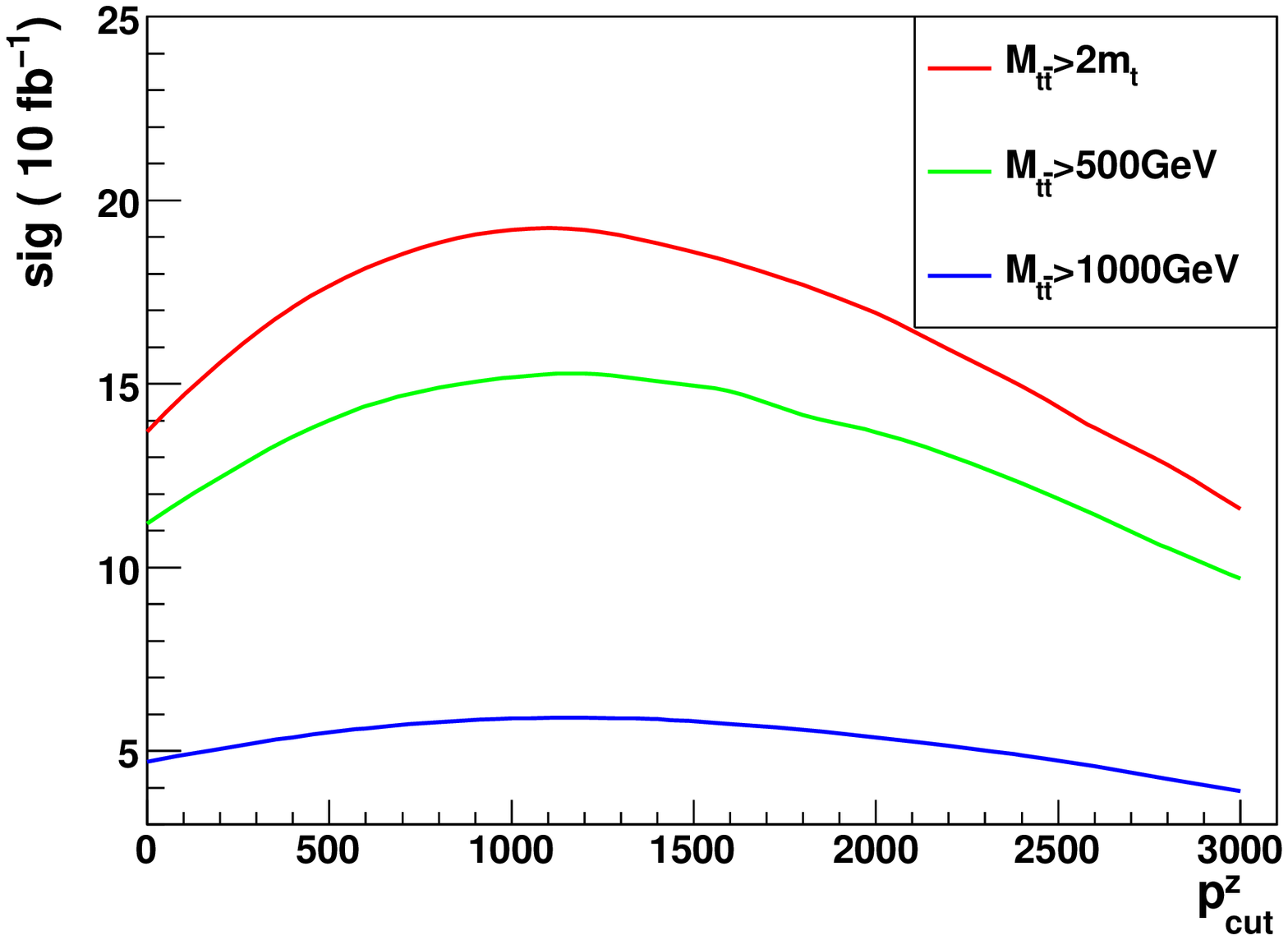} \caption{\label{Afb14TeV}
$A_{\rm OFB}$ and significance as a function of $P_{cut}^z$ at LHC
with $\sqrt{s}=14$ TeV.}
\end{figure}

Figures~\ref{Afb7TeV} and \ref{Afb14TeV} show our predictions on
$A_{\rm OFB}$ and significance to discover the asymmetric events
which is defined as ${\rm sig} =\sqrt{\cal{L}}
\sigma^A/\sqrt{\sigma} $ ( with ${\cal L}=10 {\rm fb}^{-1}$) at LHC
with $\sqrt{s}=7$ TeV and 14 TeV, respectively. For no cuts at all
($M_{t\bar{t}}>2m_t, P_{\rm cut}^z=0$), the $A_{\rm OFB}$ is not
zero but very small, $1.2\%$ and $0.58\%$ for $\sqrt{s}=7$ TeV and
14 TeV, respectively. The reason is simply due to the large
denominator $\sigma$ which arises mainly from $gg\to t\bar{t}$. To
increase $A_{\rm OFB}$, we apply cuts on $P_{t\bar{t}}^z$ and
$M_{t\bar{t}}$. The key point is to suppress $gg$ fusion
contributions to the denominator while decreasing the numerator
$\sigma^{A}$ as small as possible. From the figures, we can see
clearly that $P_{t\bar{t}}^z$ cut can increase $A_{\rm FB}$ greatly
while $M_{t\bar{t}}$ cut is not so efficient. The $P_{t\bar{t}}^z$
cut has two impacts on $A_{\rm OFB}$. First, as symmetric events
$gg\to t\bar{t}$ lie mainly in the small $P_{t\bar{t}}^z$ region,
the $P_{t\bar{t}}^z$ cut can remove them effectively. Second, as
mentioned above, it is most probably that the valence quark's
momentum is larger than that of the sea quark, but it does has some
small probability that the valence quark's momentum is smaller than
that of the sea quark. This will cause an opposite contributions to
the asymmetric cross section in our definition of $A_{\rm OFB}$. The
$P_{t\bar{t}}^z$ cut can reduce such kind of pollution.   In figures
we also show the significance to discover the asymmetric events from
the background symmetric events. Such measure can be utilized to
optimize the cut conditions, namely $M_{t\bar{t}}>2m_t$ $P_{\rm
cut}^z\sim500$ GeV is the optimal one at $\sqrt{s}=7$ TeV and
$M_{t\bar{t}}>2m_t$ $P_{cut}^z\sim1.2$ TeV is the optimal one at
$\sqrt{s}=14$ TeV though $A_{\rm OFB}$ is not the largest here. Note
that ${\rm sig} \propto \sqrt{\cal{L}}$, here we choose
$\cal{L}=10{\rm fb}^{-1}$. The value of ${\rm sig}$ varies with the
integrated luminosity but optimal cut criterion should be stable.

It's interesting to compare $A_{\rm OFB}$ proposed in this paper
with the central charge asymmetry $A_C$ proposed before [cf.
Eq.(\ref{ACharge})]. First, $A_C$ only accounts for the single top
or antitop in $|Y|<Y_C$ regions, while the new asymmetry $A_{\rm
OFB}$ needs to include the top pair kinematical information for
every events. Second, $A_C$ is based on a central region $|Y|<Y_C$,
while $A_{\rm OFB}$ is defined in a region $|P_{t\bar{t}}^z|>P_{\rm
cut}^z$. Because of the different momentum distribution of the
quarks and the corresponding sea quarks in the proton,  the top pair
events produced via $q\bar q$ annihilation are very likely be
boosted in the z direction. For the central charge asymmetry, a
small $Y_C$ cut cannot cover these $z$ direction boosted events.
However, a large $Y_C$ cut will include nearly equal number of $t$
and $\bar{t}$ events, which makes $A_C$ approach zero
\cite{Ferrario:2008wm}. $A_C$ vanishes if the whole rapidity
spectrum is integrated. Optimal $Y_C$ for $A_C$ is about 1. Our
$A_{\rm OFB}$ can cover the top pair events that reach to the edge
of the detector, so it can have more statistics. Third, the
denominator $\sigma$ in both cases is mainly composed of the
symmetric $gg\to t\bar{t}$ events. This makes both asymmetries
small, especially for $A_C$ \cite{Ferrario:2008wm}. A kinematic
feature of $gg\to t\bar{t}$ events is that they are mostly located
in a small $P_{t\bar{t}}^z$ region. So by requiring a higher $P_{\rm
cut}^z$,  top pair events via $gg$ fusion can be removed
efficiently. The central charge asymmetry does not take this
advantage so $A_C$ is smaller than $A_{\rm OFB}$.

To summarize, both  CDF and D0 at Tevatron reported the measurements
of forward-backward asymmetry in top pair production. Theoretically
such asymmetry is due to the higher-order QCD processes. The
measurements showed a possible deviation from the theoretical
prediction. In this paper, we explore how to examine the {\em same}
higher-order QCD effects at the more powerful LHC. Unlike Tevatron,
the proton-proton LHC has no preferred direction in the laboratory
frame. Thus we define a new one-side forward-backward asymmetry
$A_{\rm OFB}$ [cf. Eqs.(\ref{AFB}) and (\ref{AFB2})] in terms of the
top pair kinematical information. Our studies show that the cut on
top pair momentum in the $z$ direction can increase asymmetry
greatly. Provided that huge $t\bar t$ events will be produced at the
LHC, $A_{\rm OFB}$ can be precisely measured and compared with the
corresponding measurements at the Tevatron.

{\em Acknowledgements:} This work was supported in part by the
Natural Sciences Foundation of China (No. 10775001, No. 10635030 and
No. 11075003).

\end{document}